\documentstyle[prl,multicol,aps,epsfig]{revtex}
\newcommand \beq{\begin{eqnarray}}
\newcommand \eeq{\end{eqnarray}}
\def\simge{\mathrel{%
       \rlap{\raise 0.511ex \hbox{$>$}}{\lower 0.511ex \hbox{$\sim$}}}}
\def\simle{\mathrel{
       \rlap{\raise 0.511ex \hbox{$<$}}{\lower 0.511ex \hbox{$\sim$}}}}

\begin{document}

\title{Vortex lattices in rapidly rotating Bose-Einstein condensates: modes
and correlation functions}
\author{Gordon Baym}
\address{Department of Physics, University of Illinois at
Urbana-Champaign, 1110 West Green Street, Urbana, Illinois 61801}
\date{\today}
\maketitle

\begin{abstract}

    After delineating the physical regimes which vortex lattices encounter in
rotating Bose-Einstein condensates as the rotation rate, $\Omega$, increases,
we derive the normal modes of the vortex lattice in two dimensions at zero
temperature.  Taking into account effects of the finite compressibility, we
find an inertial mode of frequency $\ge 2\Omega$, and a primarily transverse
Tkachenko mode, whose frequency goes from being linear in the wave vector in
the slowly rotating regime, where $\Omega$ is small compared with the lowest
compressional mode frequency, to quadratic in the wave vector in the opposite
limit.  We calculate the correlation functions of vortex displacements and
phase, density and superfluid velocities, and find that the zero-point
excitations of the soft quadratic Tkachenko modes lead in a large system to a
loss of long range phase correlations, growing logarithmically with distance,
and hence lead to a fragmented state at zero temperature.  The vortex
positional ordering is preserved at zero temperature, but the thermally
excited Tkachenko modes cause the relative positional fluctuations to grow
logarithmically with separation at finite temperature.  The superfluid
density, defined in terms of the transverse velocity autocorrelation function,
vanishes at all temperatures.  Finally we construct the long wavelength single
particle Green's function in the rotating system and calculate the condensate
depletion as a function of temperature.

\end{abstract}


\section{Introduction}

    Under rapid rotation, a superfluid forms a triangular lattice of quantized
vortices carrying the angular momentum of the system.  Such structure was seen
experimentally in superfluid helium in Refs.~\cite{packard}.  Rapidly rotating
Bose-Einstein condensates of cold atoms
\cite{Madison,VortexLatticeBEC,HaljanCornell} open up the study of the physics
of vortex lattices in regimes well beyond those achievable in superfluid
helium \cite{jilatk,jila3,dalibard}.  With increasing rotational frequency,
$\Omega$, condensates go through a number of different regimes.  At $\Omega$
small compared with the lowest compressional frequencies, $\Omega \ll sk_0$,
where $s$ is the sound velocity and $k_0\sim \pi/R_\perp$ is the lowest
wavenumber in the finite geometry, with $R_\perp$ the size of the system
perpendicular to the rotation axis, the system is in the "stiff" Thomas-Fermi
regime, and responds to rotation effectively as an incompressible fluid.  As
$\Omega$ becomes larger than the lowest sound mode frequencies, $\Omega \simge
sk_0$, or essentially that the outer edge of the cloud moves supersonically,
the system enters the "soft" Thomas-Fermi regime, where the compressibility
becomes important in the response to rotation.

    When, in harmonically trapped condensates, $\Omega$ approaches the
transverse trapping frequency, $\omega_\perp$, the centrifugal force begins to
balance the trapping force, and the system flattens towards a lower density
and therefore more compressible, two-dimensional cloud.  For $\Omega$ $\gg
ms$, where $m$ is the particle mass, the condensate wave function becomes
formed primarily of particle orbits in the lowest Landau level of the Coriolis
force -- the mean field quantum Hall state \cite{Ho} -- with approximate order
parameter, for rotation about the $z$ axis,
\beq
 \Psi(r) = {\cal C}\Pi_j(u-u_j)e^{-|u|^2/2d_\perp^2},
 \quad\quad u = x+iy,
 \label{laugh}
\eeq
where the $u_j = x_j + iy_j$ are the vortex locations in the usual complex
notation, $d_\perp=(m\omega_\perp)^{-1/2}$ is the transverse oscillator
length, and $\cal C$ is a normalization constant.  The structure of the
trapped condensate in the axial direction approaches a Gaussian shape as
$|Omega\to\omega_\perp$.  Although Eq.~(\ref{laugh}) predicts that the
transverse structure is also Gaussian, the transverse structure, if
Thomas-Fermi in the non-rotating cloud, remains Thomas-Fermi in this limit
\cite{coresize}, owing to a small admixture of higher Landau levels in the
order parameter.  Experiments in this regimes are reported in
Ref.~\cite{jila3}.  With further increase of $\Omega$, the vortex lattice is
expected to melt, at the point where the number of vortices becomes of order
ten percent of the number of particles, \cite{RS,SHM}, and the system
eventually enters new highly correlated bosonic quantum Hall many-particle
states, no longer describable in mean field
\cite{Cooper,Viefers,Jolicoeur,Read,cazalilla}.

    Prior to discussing the modes of the lattice, it is useful to lay out the
demarcations between the various regimes in a harmonically trapped gas.  We
assume that the interaction between the particles, of total number $N$, is
described by a repulsive s-wave interaction parameter $g=4\pi a_s/m$, where
$a_s$ is the scattering length; we use units in which $\hbar = 1$.  In
Thomas-Fermi, the transverse radius, $R_\perp$, is given by \cite{FG},
\beq
   R_\perp^2 = \frac{d_\perp^2\tau}
   {\left(1-(\Omega/\omega_\perp)^2\right)^{3/5}},
\eeq
where $\tau = [(15Nba_s/d_\perp)(\omega_z/\omega_\perp)]^{2/5}$, and
$\omega_z$ is the axial trapping frequency.  The factor $b$, $\simge 1$,
describes the renormalization of the interaction energy of long wavelength
density fluctuations in the system \cite{b}. Furthermore,
\beq
  \frac{\Omega}{ms(0)^2}= \frac{2}{\tau}\frac{\Omega}{\omega_\perp}
  {(1-(\Omega/\omega_\perp)^2)^{2/5}},
\eeq
where $s(0)$ is the sound velocity in the center of the cloud.  If we
write $k_0 = \alpha/R_\perp$, where $\alpha \simeq 5.45$ \cite{crescimanno},
the criterion for being in the soft regime, that $\Omega/sk_0$ be large,
becomes
\beq
   \frac{\Omega/\omega_\perp}{\left(1-(\Omega/\omega_\perp)^2\right)^{1/2}}
     \gg \frac{1}{\sqrt2 \alpha} \simeq 0.13.
\eeq
The experiments of Ref~\cite{jilatk} on Tkachenko modes reach $\Omega/sk_0
\sim 1.15$.

    The criterion to be in the mean field quantum Hall regime, $\Omega/ms^2
\gg 1$, can be conveniently written in terms of the filling factor, $\nu =
N/N_v$, where $N_v$ is the total number of vortices in the rotating cloud.
Since
\beq
  N_v = \pi n_v R_\perp^2 = m\Omega R_\perp^2 =
  \frac{\Omega}{\omega_\perp}\frac{\tau}
  {(1-(\Omega/\omega_\perp)^2)^{3/5}},
\eeq
where $n_v = m\Omega/\pi$ is the density of vortices, we find
\beq
  \frac{\Omega}{ms(0)^2}=
  \frac{2}{\nu^{2/3}}\left(\frac{\Omega}{\omega_\perp}\right)^{1/3}
  \left(\frac{d_\perp\omega_\perp}{15ba_s\omega_z}\right)^{2/3}.
\eeq
This result is independent of the number of particles, as long as the
system is in the Thomas-Fermi regime.  For the parameters
$\omega_\perp/\omega_z = 8.3/5.2$ of the Tkachenko mode experiment of
Ref.~\cite{jilatk} in $^{87}$Rb, we find $\Omega/ms(0)^2 \simeq 200/\nu^{2/3}$
in the limit $\Omega \to \omega_\perp$.

    The vortex lattice supports a number of modes, first discussed by
Tkachenko \cite{Tkachenko} for a two-dimensional incompressible fluid, and
later in Ref.~\cite{BC} at finite temperature with full effects of the normal
fluid, dissipation and Kelvin oscillations of the vortex lines in three
dimensions.  The low frequency in-plane Tkachenko mode is an elliptically
polarized oscillation of the vortex lines, with the semi-major axis of the
ellipse orthogonal to the direction of propagation.  The Tkachenko mode is
linear at small wave vector, $k$, in the transverse plane,
\beq
 \omega_T = \left(\frac{2C_2}{mn}\right)^{1/2}k \quad \to
  \left(\frac{\Omega}{4m}\right)^{1/2}k,
\label{omegatk}
\eeq
where $n$ is the particle density, and $C_2$ is the elastic shear modulus
of the vortex lattice; at slow rotation, $C_2=n\Omega/8$.  In the soft regime,
the dispersion relation instead becomes quadratic \cite{SHM,tkmodes},
\beq
 \omega_T = \left(\frac{s^2C_2}{2\Omega^2 nm}\right)^{1/2}k^2.
\eeq
Were it possible to rotate helium sufficiently rapidly one would also see
this very long wavelength quadratic behavior.

    Such Tkachenko soft modes can play havoc with the stability of a large
system, causing loss of long range phase coherence even at zero temperature;
they are eventually responsible for the melting of the lattice \cite{SHM}.  In
a recent paper \cite{tkmodes} we derived the modes of the vortex lattice for
all rotation rates, through constructing the conservation laws and superfluid
acceleration equation describing the long wavelength behavior of the system.
In this paper we focus on deriving the correlation functions of density,
superfluid phase and velocity, and vortex displacements from equilibrium,
which enable us to understand the effects of the soft infrared structure on
the stability of the superfluid and lattice.  This work is a generalization of
Ref.~\cite{lattice}, which discussed the effects of the oscillations of the
vortex lines at finite temperature in liquid helium on the long ranged phase
correlations of the superfluid.  As we shall see, the long wavelength
Tkachenko modes lead to fragmentation of the condensate, even at zero
temperature.  Whether the system loses phase coherence over its volume or the
lattice melts first depends on the number of particles in the system and its
rotation rate.

    In Sec. II, we review the basic equations describing the dynamics,
restricting the analysis to linearized motion in two dimensions, and
neglecting the normal component of the superfluid as well as dissipative
terms.  The analysis of the full three-dimensional problem will be published
separately \cite{drew}.  In Sec. III we construct the correlation functions of
the physical quantities of interest, and in Sec. IV study the condensate
depletion by constructing the single particle Green's function for the
rotating superfluid.

\section{Condensate phase and conservation laws}

    Let us first recapitulate the conservation laws and equation for the
superfluid phase which govern the long wavelength behavior of the system.
While much of this material has been given in Ref.~\cite{tkmodes}, we include
it here in order to facilitate the derivation of the correlation functions.
The basic formalism given here applies to general bosonic superfluids.

\subsection{Condensate phase}

    We work in the frame co-rotating with the lattice, and describe the
deviations of the vortices from their home positions by the continuum
displacement field, $\epsilon(r,t)$.  In linear order in the vortex
displacements, the long wavelength superfluid velocity, $\vec v(r,t)$, can be
written, following Ref.~\cite{lattice}, in terms of the long wavelength
vortex-lattice displacement field and the phase $\Phi(r,t)$ of the order parameter,
as
\beq
   \vec v + 2\vec\Omega\times\vec\epsilon = {\vec \nabla} \Phi/m;
 \label{vphase}
\eeq
The curl of this equation is
\beq
   {\vec \nabla} \times \vec v = -2\Omega {\vec \nabla} \cdot \vec\epsilon.
  \label{curlv}
\eeq
The origin of Eq. (\ref{vphase}) is the law of conservation of vorticity,
$\vec\varpi \equiv {\vec \nabla} \times \vec v$:
\beq
   \frac{\partial \vec\varpi}{\partial t} + {\vec \nabla} \times (\vec\varpi \times
    \dot{\vec\epsilon}) = 0,
\eeq
where here $\dot{\vec\epsilon}$ tells the rate at which the vorticity
moves about.  Since under uniform rotation, $\vec\varpi = 2\vec\Omega$, the
time derivative of the curl of Eq.~(\ref{vphase}) is just this equation
linearized.  The longitudinal part of the left side of Eq.~(\ref{vphase}) is
trivially the gradient of a scalar.  Equation~(\ref{vphase}) constrains the
number of degrees of freedom in two dimensions to four from the original five
-- $n$, $\vec v$, and $\vec \epsilon$.

    The time derivative of Eq.~(\ref{vphase}) is the superfluid acceleration
equation,
\beq
  m\left({{\partial \vec v}\over{\partial t}} + 2\vec\Omega \,\times
    \dot{\vec
    \epsilon}\right) = - {\vec \nabla} (\mu - V_{\rm eff}),
  \label{supaccel}
\eeq
where $\mu$ is the chemical potential.  For an axially symmetric harmonic
confining trap of frequency $\omega_\perp$ in the transverse direction and
$\omega_z$ in the axial direction,
\beq
 V_{\rm eff} = \frac m2 \left[(\omega_\perp^2-\Omega^2) r^2 + \omega_z^2
   z^2\right],
\eeq
where $\vec r\,$ denotes $(x,y)$.  In the frame corotating with the vortex
lattice, the chemical potential $\mu$ is related to the phase by
\beq
  \mu(r,t) - V_{\rm eff} = -\frac{\hbar}{m}
   \frac{\partial \Phi(r,t)}{\partial t}.
  \label{mu-phi}
\eeq

\subsection{Conservation laws}

    The dynamics are specified by the conservation laws of particles and
momentum, together with the superfluid acceleration equation,
(\ref{supaccel}).  The continuity equation takes its usual form,
\beq
  \frac{\partial n(r,t)}{\partial t} + {\vec \nabla} \cdot \vec j(r,t) = 0,
  \label{contin}
\eeq
where $n$ is the density and $\vec j=n\vec v$ the particle current.
Conservation of momentum reads:
\beq
   m\frac{\partial \vec j}{\partial t} +  2m\vec\Omega\times \vec
  j +{\vec \nabla} P + n{\vec \nabla}  V_{\rm eff} = -\vec \sigma -\vec \zeta.
  \label{momcons}
\eeq
Here $P$ is the pressure, and $\vec\sigma$ is the elastic stress tensor,
discussed below.  At zero temperature, ${\vec \nabla} P = n{\vec \nabla} \mu$,
while in equilibrium, ${\vec \nabla} P +n{\vec \nabla} V_{\rm eff} = 0$.  To
calculate the displacement autocorrelation functions, we include here, as
in \cite{lattice}, an external driving force, $-\vec\zeta\,(\vec r, t)$,
acting on the lattice, derived from an external perturbation
$H'=\vec\zeta\cdot\vec\epsilon$.

    The elastic stress tensor is derived from the elastic energy density of
the lattice, which in two dimensions has the form (in the notation of
\cite{BC}),
\beq
 {\cal E}(r) = 2C_1 ({\vec \nabla}\cdot\vec\epsilon\,)^2
        +C_2\left[\left(\frac{\partial \epsilon_x}{\partial x}
          -\frac{\partial\epsilon_y}{\partial y}\right)^2
    + \left(\frac{\partial \epsilon_x}{\partial y}  +\frac{\partial
     \epsilon_y}{\partial x}\right)^2\right],
 \label{elastic}
\eeq
where $C_1$ is the compressional modulus, and $C_2$ the shear modulus of
the vortex lattice.  The elastic constants are density-dependent properties of
the fluid.  Then the elastic stress tensor, $\sigma_i$, is given in terms of
the total elastic energy, $E_{\rm el} = \int d^2r {\cal E}(r)$, by
\beq
 \sigma_i(r,t) = \frac{\delta E_{\rm el}}{\delta \epsilon_i}
               = -4\nabla_i \left(C_1{\vec \nabla}\cdot\vec\epsilon\,\right)
                 -2{\vec \nabla}\cdot\left(C_2{\vec \nabla}\epsilon_i\right),
 \label{sigi}
\eeq
where we allow for effects of density gradients entering through the
elastic constants.  In an incompressible fluid, $C_2 = n\Omega/8 = -C_1$.  On
the other hand, in the quantum Hall regime, the shear modulus is determined by
the deviations of the interaction energy caused by distorting the vortex
lattice, and \cite{elastic}
\beq
  C_2  \simeq \frac{81}{80\pi^4}ms^2n,
 \label{C2}
\eeq
in agreement with the shear modulus given numerically in \cite{SHM}.
Calculation of the elastic constants $C_1$ and $C_2$ over the full range of
$\Omega$ from the stiff Thomas-Fermi to the quantum Hall limits will be given
in Ref.~\cite{elastic}.

    Subtraction of Eq.~(\ref{momcons}) divided by $n$, from the superfluid
acceleration equation (\ref{supaccel}), with ${\vec \nabla} P = n{\vec \nabla}\mu$, yields
\beq
  2\vec \Omega\times(\dot{ \vec\epsilon}-\vec v\,) =
      \frac{\vec \sigma +\vec \zeta}{mn}.
 \label{epsv}
\eeq
Equations~(\ref{supaccel}), (\ref{contin}), (\ref{sigi}), and (\ref{epsv})
fully specify the problem for a trapped system with a non-uniform density.

\subsection{Modes}

    In the following, we neglect effects of non-uniformity of the equilibrium
density for simplicity, and proceed to derive the modes as in \cite{tkmodes}.
The curl of Eq.~(\ref{epsv}) is
\beq
   {\vec \nabla}\cdot ({\dot{\vec\epsilon}} - \vec v\,)
   = \frac{1}{2\Omega nm}{\vec \nabla}\times(\vec \sigma+\vec \zeta\,),
 \label{diveps}
\eeq
while its divergence, together with (\ref{curlv}), yields,
\beq
  {\vec \nabla}\times {\dot {\vec \epsilon}}\, +
         2\vec\Omega \, {\vec \nabla} \cdot{\vec\epsilon}
             = -\frac{1}{2\Omega nm}{\vec \nabla}\cdot(\vec\sigma+\vec\zeta\,),
 \label{xeps}
\eeq
where ${\vec \nabla}\times\vec\sigma = -2C_2\nabla^2({\vec
\nabla}\times\vec\epsilon)$, and ${\vec \nabla}\cdot\vec\sigma =
-2(C_2+2C_1)\nabla^2({\vec \nabla}\cdot\vec\epsilon)$.  To eliminate $\vec v$
from (\ref{diveps}), we note that the density oscillations are governed by
\beq
 \left(-\frac{\partial^2}{\partial t^2} + s^2\nabla^2\right) n =
             2n\Omega{\vec \nabla}\times{\dot \epsilon},
 \label{densosc}
\eeq
where the sound speed, $s$, is given by $ms^2=\partial P/\partial n$
\cite{sound}.  In terms of the frequency, $\omega$, and wave vector, $k$, we
have then
\beq
    \vec k\cdot \vec v = \frac{2\omega^2}{\omega^2-s^2k^2}\vec \Omega \cdot
  \vec k \times \epsilon,
 \label{vepst}
\eeq
so that in terms of longitudinal and transverse components \cite{notation},
\beq
    -i\omega \epsilon_T +\left(2\Omega+\frac{C_2+2C_1}{\Omega nm}k^2
    \right)\epsilon_L = -\frac{\zeta_L}{2\Omega nm}; \nonumber \\
    -i\omega \epsilon_L -\left(\frac{2\omega^2\Omega}{\omega^2-s^2k^2}
       +\frac{C_2}{\Omega mn}k^2\right) \epsilon_T
        = \frac{\zeta_T}{2\Omega nm}.
  \label{epstl}
\eeq
Solving for $\epsilon_L$ and $\epsilon_T$ we have
\beq
    \epsilon_L &=&  \frac{1}{nmD}\left\{
     \left(\omega^2 + (\omega^2-s^2k^2)\frac{C_2}{2\Omega^2
   nm}k^2\right)\zeta_L +
          i\omega\frac{\omega^2-s^2k^2}{2\Omega}\zeta_T\right\},
    \nonumber \\
 \epsilon_T &=&  \frac{1}{nmD}\left\{
     (\omega^2-s^2k^2)\left(1+\frac{C_2+2C_1}{2\Omega^2 nm}k^2\right)\zeta_T
          -i\omega\frac{\omega^2-s^2k^2}{2\Omega}\zeta_L\right\},
\eeq
where the secular determinant, whose zeroes determine the mode frequencies, is
\beq
  D(k,\omega) \equiv \omega^4 - \omega^2\left[4\Omega^2 + \left(s^2
   +\frac{4}{nm}(C_1+C_2)\right)k^2\right]+ \frac{2s^2C_2}{nm}k^4 =
   (\omega^2 -\omega_I^2)(\omega^2 -\omega_T^2) =0;
 \label{D}
\eeq
we have dropped terms of second order in the elastic constants.

    For $2s^2C_2k^4/nm \ll (4\Omega^2 + (s^2 +4(C_1+C_2)/nm)k^2)^2$, as is
always the case at long wavelengths in both the incompressible and quantum
Hall limits, the mode frequencies are given by
\beq
  \omega_I^2 = 4\Omega^2 +
  \left(s^2+\frac{4(C_1+C_2)}{nm}\right)k^2,
\eeq
and
\beq
\omega_{T}^2 =
  \frac{2C_2}{nm} \frac{s^2k^4}{(4\Omega^2 + (s^2+4(C_1+C_2)/nm)k^2)}.
\label{tk}
\eeq
The first mode is the standard inertial mode of a rotating fluid; for
$\Omega \ll s^2k^2$ the mode is a sound wave, while for $\Omega \gg
s^2k^2$, the mode frequencies begin essentially at twice the axial trapping
frequency.  This mode has been calculated in realistic trapping geometries in
\cite{cozzini} and \cite{bigelow}.  The second mode is the elliptically
polarized Tkachenko mode.  Equation~(\ref{xeps}) implies that the inertial
mode is circularly polarized:  $\epsilon_L/\epsilon_T \simeq i$, and in the
Tkachenko mode, $\epsilon_L/\epsilon_T \simeq i\omega_T/2\Omega$; the small
longitudinally polarized component is $\pi/2$ out of phase with the
transversely polarized component.  In the limit of an incompressible fluid
($s^2\to\infty$),
\beq
  \omega_I^2 = 4\Omega^2 + \left(s^2+\frac{4C_2}{nm}\right)k^2,
\eeq
and
the Tkachenko frequency, $\omega_T$, is linear in $k$,
Eq.~(\ref{omegatk}).
 In the soft limit, by contrast,
\beq
  \omega_T^2 = \frac{s^2C_2}{2\Omega^2 nm}k^4;
\eeq
unlike in the stiff Thomas-Fermi regime, the mode frequency is
quadratic in $k$ at long wavelengths; using Eq.~(\ref{C2}) for $C_2$ we have
\beq
  \omega_T \simeq \frac{9}{4\pi^2\sqrt{10}}\frac{s^2k^2}{\Omega}.
 \label{softtk}
\eeq
The present results for the modes are valid for a uniform system over the
entire range of rotation frequencies, from the slowly rotating stiff regime up
to the melting of the vortex lattice.  The Tkachenko mode has been calculated
numerically for realistic trapping geometries in \cite{crescimanno} in
the stiff limit and more generally in \cite{bigelow2}.

\section{Correlation functions}

    We turn to determining the effects of the lattice modes on the lattice
ordering, and the phase coherence and condensate fraction of the rotating
superfluid.  To do so we construct the correlation functions of the density,
superfluid velocity, vortex displacements, and phase from the dynamical
equations in the previous section.  All the correlation functions of interest,
including the single particle Green's function, can be written in terms of the
density-density and displacement-displacement correlation functions.  In the
following we let $\langle AB\rangle(k,z)$ denote the Fourier transform in
space and the analytic continuation of the Fourier transformation in imaginary
time to complex frequency, $z$, of the correlation function $\langle A(rt)
B(r't')\rangle - \langle A(rt)\rangle \langle B(r't')\rangle$.

    The density-density correlation function, $\langle nn\rangle(k,z)$,
is readily found from the response of $\langle n(r,t)\rangle$ to an external
potential, $U(r,t)$, coupled to the density.  Using Eq.~(\ref{vepst}) to
eliminate $\epsilon_T$ from (\ref{densosc}), we have
\beq
   \langle nn\rangle(k,z) =  \int_{-\infty}^\infty \frac{d\omega}{2\pi}
           \frac{B(b,\omega)}{z-\omega} =
    \frac{nk^2}{mD(k,z)}\left(z^2-\frac{2C_2k^2}{nm}\right),
 \label{nncorr}
\eeq
with $D$ given by (\ref{D}).  As $z\to\infty$, $\langle
nn\rangle(k,z)\to nk^2/mz^2$, which is the expected f-sum rule on
the spectral weight $B(k,\omega)$,
\beq
    \int_{-\infty}^\infty \frac{d\omega}{2\pi} \omega B(k,\omega) =
    \frac{nk^2}{m}.
\eeq
Similarly, $\langle nn\rangle(k\to 0,0)\to -n/ms^2$, yielding the correct
compressibility sum rule,
\beq
  \lim_{k\to 0} \int_{-\infty}^\infty
   \frac{d\omega}{2\pi}\frac{B(b,\omega)}{\omega} =  \frac{n}{ms^2}.
\eeq
As expected, the f-sum rule is dominated by the high frequency inertial
mode; the low frequency Tkachenko mode dominates the compressibility sum rule
\cite{cps}.

    The correlations of the longitudinal velocity are given in terms of
$\langle nn\rangle(k,\omega)$, as usual, by
\beq
  \langle v_L n\rangle(k,z) =
  \frac{z}{nk}\langle n n\rangle(k,z),
 \label{vn}
\eeq
and
\beq
  \langle v_L v_L\rangle(k,z) &=&
  \frac{z^2}{n^2k^2}\langle nn\rangle(k,z) -  \frac{1}{nm}
\\
 &=&\frac{1}{nmD}\left\{z^2\left(4\Omega^2 + s^2k^2
   +\frac{2}{nm}(2C_1+C_2)k^2\right) - \frac{2s^2C_2}{nm}k^4 \right\}.
\label{vv}
\eeq

    The correlation functions of the elastic displacements are given
by
\beq
  \langle \epsilon_i \epsilon_j \rangle
  = \int_{-\infty}^{\infty}
  \frac{d\omega}{2\pi} \frac{B_{ij}(k,\omega)}{z-\omega}
  = \frac{\delta \langle \epsilon_i\rangle}{\delta \zeta_j}.
 \label{deltaz}
\eeq
where $\langle\epsilon_i\rangle$ is the displacement in the $i^{\rm th}$
direction induced by the force $\vec\zeta$.  Equations~(\ref{deltaz}) and
(\ref{epstl}) then imply,
\beq
    \langle\epsilon_L\epsilon_L\rangle(k,z) &=&  \frac{1}{nmD(k,z)}
     \left(z^2 + (z^2-s^2k^2)\frac{C_2}{2\Omega^2 nm}k^2\right)
    \label{longcorr}
     \\
    \langle\epsilon_T\epsilon_T\rangle(k,z) &=& \frac{z^2-s^2k^2}{nmD(k,z)}
    \left(1+\frac{C_2+2C_1}{2\Omega^2 nm}k^2\right)
   \label{transcorr}
     \\
    \langle\epsilon_L\epsilon_T\rangle(k,z) &=&
          iz\frac{z^2-s^2k^2}{2\Omega nmD} =
           \langle\epsilon_T\epsilon_L\rangle^*.
    \label{longtran}
\eeq

    We note, for later calculation of the phase correlations, that the
displacement-density correlation function, found from Eq.~(\ref{densosc})
together with the continuity equation, is
\beq
 \langle n\epsilon_T \rangle(k,z) =  \langle\epsilon_T n
 \rangle(k,z)  =  \frac{2n\Omega z k}{z^2-s^2k^2}
 \langle\epsilon_T\epsilon_T\rangle,
 \label{neps}
\eeq
and thus the displacement-longitudinal velocity correlation is,
\beq
  \langle\epsilon_T v_L \rangle(k,z)  = \langle v_L\epsilon_T
 \rangle(k,z)  =
   \frac{2\Omega z^2}{z^2-s^2k^2}
\langle\epsilon_T\epsilon_T\rangle.
 \label{vep}
\eeq

\subsection{Lattice displacements}

    We first address the effects of the vortex modes on the lattice
displacements from equilibrium.  At finite temperature, $T$, the equal time
displacement correlations are given by
\beq
   \langle\left(\epsilon_i(r)-\epsilon_i(r')\right)^2\rangle
      = 2\int \frac{d^2 k}{(2\pi)^2Z}\left(1-\cos \vec k\cdot \vec R\right)
      \int_0^\infty \frac{d\omega}{2\pi}B_{ii}(k,\omega)(1+2f(\omega)),
\eeq
where $Z$ is the thickness of the system in the z direction, $\vec R =
\vec r-\vec r\,'$, and $f(\omega) = 1/(e^{\beta\omega}-1)$, with $\beta =
1/k_B T$.  The spectral weights $B_{ij}$ are found from
Eqs.~(\ref{longcorr})-(\ref{longtran}) by letting $z\to\omega$ and
\beq
    \frac{1}{D(k,z)} \to
 \frac{2\pi}{\omega_I^2-\omega_T^2}\left\{\frac{1}{2\omega_I}
 \left(\delta(\omega -\omega_I)- \delta(\omega -\omega_I)\right)
  - \frac{1}{2\omega_T}
 \left(\delta(\omega -\omega_T)- \delta(\omega -\omega_T)\right) \right\}.
\eeq

    The leading terms in the mean displacement of a single vortex from
equilibrium due to excitations of the modes are
\beq
   \langle{\vec\epsilon\,}^2\rangle
      = \int \frac{d^2 k}{(2\pi)^2Z}
       \frac{1}{\omega_I^2 nm}
         \left[\omega_I(1+2f(\omega_I)) +
         \frac{s^2k^2}{2\omega_T}(1+2f(\omega_T))\right].
 \label{meandis}
\eeq
The mean displacement is convergent at zero temperature.  However, at
finite temperature it diverges logarithmically with system size if the
Tkachenko mode spectrum reaches down into the soft quadratic regime; then
\beq
   \frac{ \langle\vec\epsilon\,^2\rangle }{\ell^2}
  \sim \frac{Tm\Omega}{8\pi ZC_2} \ln N_v.
\eeq
The relative separation, with only the leading terms kept,
\beq
   \langle\left(\vec\epsilon\,(r)-\vec\epsilon\,(r')\right)^2\rangle
      = 2\int \frac{d^2 k}{(2\pi)^2Z}
       \frac{1-\cos \vec k\cdot \vec R}{nm\omega_I^2}
         \left[\omega_I\left(1+2f(\omega_I)\right)
        + \frac{s^2k^2 }{2\omega_T} \left(1+2f(\omega_T)\right) \right],
\eeq
converges at zero temperature, and lattice preserves long range positional
order at large separation; at finite temperature, however,
$\langle\left(\vec\epsilon\,(r)-\vec\epsilon\,(r')\right)^2\rangle$ grows
logarithmically with separation, $\sim (Tm\Omega/4\pi ZC_2) \ln N_v(R)$, where
$N_v(R)$ is the number of vortices within radius $R$.  In the stiff
Thomas-Fermi limit, this expression becomes $(4\pi/nZ\lambda^2)\ln N_v(R)$,
where $\lambda$ is the thermal wavelength; in the quantum Hall limit we have
rather $(20\pi^4/81\nu)(T/ms^2) \ln N_v(R)$.

    Equation~(\ref{meandis}) may be used with a Lindemann criterion to
estimate the point where the lattice melts at $T=0$ in an extended system
\cite{RS,SHM}.  The zero temperature displacements are sensitive to the entire
spectrum of modes up to the lattice Debye vector, $k_d = (4\pi n_v)^{1/2}$,
whereas the mode frequencies derived here are valid only for $k \ll k_d$.  For
a first estimate, we replace the integrand by its infrared limit, letting
$\int d^2k/(2\pi)^2 \to n_v$.  Then using the Tkachenko frequency at rapid
rotation, (\ref{softtk}), we have
\beq
   \frac{\langle{\vec\epsilon\,}^2\rangle}{\ell^2}
      =  \frac{1}{2\nu}
       \left(1+ \frac{s^2k^2}{4\omega_T\Omega}\right)
        \simeq \frac{1}{2\nu}\left(1+\frac{\pi^2{\sqrt10}}{9}\right)
        \simeq \frac{2.23}{\nu},
\label{eps2}
\eeq
where $\ell = (1/\pi n_v)^{1/2}$ is the radius of the Wigner-Seitz cell
around a given vortex, and we have used (\ref{C2}).  The ``1" term arises from
the inertial mode, with equal contributions from the transverse and
longitudinal displacements; the final term arises solely from the soft
Tkachenko mode contribution to the transverse displacements.  In order to take
into account approximately the mode structure at larger wave vector, we
include the kinetic term $-\nabla^2 n/4m$ in the pressure.  In the
non-rotating weakly interacting gas, this term modifies the linear spectrum,
$sk$, into the full Bogoliubov spectrum, $E_k= \left((sk)^2
+(k^2/2m)^2\right)^{1/2}$.  Inclusion of such a term is equivalent to
replacing $s^2$ by $s^2 + k^2/4m^2$ in the mode spectrum, with the effect of
stiffening the Tkachenko mode at high wave vector.  Then at $\Omega/ms^2 \sim
40$, as expected under typical experimental conditions at $\nu\sim 10$, the
contribution of the Tkachenko modes to $\langle \epsilon^2 \rangle$ is reduced
from $1.73/\nu$ in Eq.~(\ref{eps2}) to $\sim 0.72/\nu$, a result consistent
with that reported in Ref.~\cite{SHM},
$m\Omega\langle{\vec\epsilon\,}^2\rangle \simeq 0.66/\nu$.  Then with the
contribution of the inertial modes included,
\beq
   \frac{\langle{\vec\epsilon\,}^2\rangle}{\ell^2}
        \simeq \frac{1.22}{\nu}.
\label{eps3}
\eeq
Taking the Lindemann criterion for melting in two dimensions put forth in
\cite{RS}, $\langle{\vec\epsilon\,}^2\rangle/\ell^2\simeq 0.07$, we find
melting of the lattice at filling $\nu \sim 17$; were we to take only the
Tkachenko mode contribution, the melting would be at $\nu \sim 10$.

\subsection{Phase correlations}

    We next determine the effect of the vortex excitations on the correlations
of the order parameter in the superfluid and on the condensate fraction.  The
condensate density, $n_0$, is most conveniently found, \`a la Onsager-Penrose,
as the limit of the single particle density matrix for large separation:
\beq
\langle \psi(r)\psi^\dagger(r')\rangle \to n_0, \quad\quad
|\vec r-\vec r\,'|\to \infty,
\eeq
where $\psi(r)$ is the single particle annihilation operator \cite{haensch}.
We write $\psi(r)$ in terms of the density and phase operators, viz.,
$\psi(r) =\sqrt{n(r)}e^{i\Phi(r)}$, and expand the long wavelength structure
to second order in small fluctuations of $n$ about $\bar n$, its equilibrium
value, and $e^{i\Phi(r)}$ about unity.  Then in terms of equal time
correlation functions,
\beq
  \langle \psi(r)\psi^\dagger(r')\rangle &\simeq&
    {\bar n} \langle e^{i\Phi(r)} e^{-i\Phi(r')}\rangle +
   \frac{1}{4{\bar n}}
   \left(\langle \delta n(r)\delta n(r')\rangle
         -\langle (\delta n(r))^2\rangle\right)
     \nonumber \\
   && +\frac12\langle
    \delta n(r)(e^{i\Phi(r)}-e^{-i\Phi(r')})+
   (e^{i\Phi(r)}- e^{-i\Phi(r')}) \delta n(r') \rangle,
 \label{psipsi}
\eeq
where $\delta n(r) = n(r) - {\bar n}$, and $\bar n$ denotes the average
density.  The first term on the right is the U(1)-invariant correlation of the
order parameter, given, for Gaussianly-distributed Fourier components of the
phase, by
\begin{eqnarray}
  \langle e^{i\Phi(r)} e^{-i\Phi(r')}\rangle
   = e^{-{1\over2}\langle (\Phi(r)- \Phi(r'))^2\rangle}.
\label{26}
\end{eqnarray}

    Equation~(\ref{nncorr}) implies that as $|\vec r-\vec r\,'|\to\infty$,
$\langle \delta n(r) \delta n(r')\rangle$ and also the final bracket in
(\ref{psipsi}) vanish.  Thus the density of particles in the condensate is
given by,
\beq
  n_0 = \lim_{|r-r'|\to\infty}\left[
  {\bar n} e^{-\frac12\langle (\Phi(\vec r)- \Phi(\vec r\,'))^2\rangle}
   -\frac{1}{4{\bar n}}\langle \delta n(r)^2 \rangle\right].
 \label{n0}
\eeq
When the phase flucutations are convergent, and vanishing at large
separation, we can expand to second order to find the usual expression for the
condensate depletion in the Bogoliubov approximation (see Sec. IV):
\beq
  n' = {\bar n} \langle\Phi^2\rangle
   +\frac{1}{4\bar n}\langle \delta n(r)^2 \rangle.
\eeq

    To determine the phase-phase and the density-phase correlations, we note
that the divergence of Eq.~(\ref{vphase}) implies that the Fourier compenents
of the phase obey $\Phi = -(im/k)(v_L - 2\Omega\epsilon_T)$.  Thus
\beq
     \langle\Phi\Phi\rangle(k,z) &=&
    \frac{m^2}{k^2}\left[ \langle v_L v_L\rangle(k,z)
     -2\Omega\left(\langle\epsilon_T v_L \rangle(k,z) +
     \langle v_L\epsilon_T \rangle(k,z)\right)
    +4\Omega^2\langle\epsilon_T\epsilon_T\rangle(k,z)\right],
\eeq
which, with Eqs.~(\ref{vv}), (\ref{transcorr}), and (\ref{vep}), becomes,
\beq
   \langle\Phi\Phi\rangle(k,z)  &=&
   \frac{m^2}{k^2}\left[\frac{z^2}{n^2k^2}\left(\langle nn\rangle
    - \frac{nk^2}{mz^2}\right)   -
  4\Omega^2\frac{z^2+s^2k^2}{z^2-s^2k^2}
    \langle\epsilon_T\epsilon_T\rangle\right]
   \nonumber \\
   &=& \frac{ms^2}{nD}\left(z^2-4\Omega^2\
   -\frac{4(C_1+C_2)}{nm}k^2 \right).
 \label{phiphi}
\eeq
Similarly,
\beq
      \langle\Phi n\rangle(k,z) =
       -i\frac{mz}{nk^2}\left[\langle nn\rangle (k,z)
      -\frac{4\Omega^2n^2k^2}{z^2-s^2k^2}
      \langle\epsilon_T\epsilon_T\rangle\right] =
               -iz\frac{n}{ms^2}\langle\Phi\Phi\rangle(k,z).
 \label{phin}
\eeq
Note that the phase fluctuations are more divergent in the infrared limit
by a factor $1/k^2$ than the transverse displacement fluctuations.

    The relative phase correlations are given by:
\beq
   \langle\left(\Phi(r)-\Phi(r')\right)^2\rangle
      = \frac{ms^2}{n}\int \frac{d^2 k}{(2\pi)^2Z}
       \frac{1-\cos \vec k \cdot \vec R}{\omega_T} (1+2f(\omega_T)),
 \label{relphase}
\eeq
plus finite terms, where $\vec R = \vec r - \vec r'$.  In the stiff limit
($s^2k^2\gg \Omega^2$), the relative phase fluctuations are convergent at zero
temperature, but, as discussed in \cite{lattice}, the finite temperature
contribution of the Tkachenko modes washes out the phase correlations
logarithmically, and the system is in fact in a fragmented condensate phase
\cite{frag}.  One finds similar behavior at zero temperature in the soft
limit.  The zero temperature part of the integral (\ref{relphase}) is $\sim
\ln (k_D R)$.  Thus from Eq.~(\ref{26}), we find the correlation of the phase
factors,
\beq
\langle e^{i\Phi(r)}e^{-i\Phi(r')}\rangle
    \sim (k_D R)^{-\eta} \sim N_v(R)^{-\eta/2},
  \label{alg}
\eeq
where again $N_v(R)$ is the number of vortices within radius $R$, and
\beq
   \eta = \frac{1}{\nu}
   \left(\frac{ms^2{\bar n}}{8C_2}\right)^{1/2}.
\eeq
From the limits $\Omega \ll ms^2$ to $\Omega \gg ms^2$, the range of $\eta$
is,
\beq
   \frac{1}{\nu}\left(\frac{ms^2}{\Omega}\right)^{1/2} \ge \eta \ge
   \frac{\pi^2\sqrt{10}}{9\nu};
\eeq
is .  The phase
correlations fall off algebraically at large $R$, as expected for a
two-dimensional system \cite{slinky}.  As the phase correlations decrease, the
condensate fraction also falls, as $(n_0/n)\sim N_v^{-\eta/2}$.

    The falloff of the phase correlations begins to become important for
$(\eta/2)\ln N_v\simge 1$, which translates effectively into the condition in
the quantum Hall limit that $\nu \simle 1.7 \ln N_v$, or $N \simle \nu
e^{0.6\nu}$; for $N= 10^6$, the condition is that $N_v \simge 5\times 10^4$;
for $N= 10^4$, $N_v \simge 10^3$; and for $N= 10^3$, one needs only $N_v\simge
10^2$ to find loss of phase coherence across the system.

    One may ask whether the falloff is significant by the time the lattice
will have melted.  The most divergent terms in the phase fluctuations are
induced by the transverse displacement fluctuations in Eq.~(\ref{phiphi}), so
that in a system of transverse radius $R_\perp$, one has roughly,
\beq
   \lim_{|\vec r-\vec r\,'|\to R_\perp}
   \frac12\langle(\Phi(r)-\Phi(r'))^2\rangle
   \simeq \frac{8}{\ell^4}\langle k^{-2}\rangle
   \langle \epsilon_T^2\rangle ,
\eeq
where the average $\langle k^{-2}\rangle \sim (1/4\pi n_v)\ln N_v$
is taken with respect to the weight in the transverse displacement correlation
function.  Thus
\beq
   \lim_{|\vec r-\vec r'|\to R_\perp}\frac12\langle(\Phi(r)-\Phi(r'))^2\rangle
      \sim 2\frac{\langle\epsilon_T^2\rangle}{\ell^2}\ln N_v.
  \label{phieps1}
\eeq
The Lindemann criterion, $\langle\epsilon_T^2\rangle/\ell^2\sim 0.07$,
implies that for $N_v\simge 10^3$ the right side of Eq.~(\ref{phieps1})
exceeds unity at the melting point, setting the scale for number of vortices
for which loss of long range order of the condensate prior to melting
becomes important.

    At finite temperature, the phase correlation integral (\ref{relphase}) is
logarithmically singular in the infrared for all $R$, indicating that in two
dimensions, the system is, as expected, no longer Bose condensed.  The
situation is the same as in the stiff Thomas-Fermi limit, where the phase
correlations in two dimensions diverge in the infrared, but in three
dimensions fall algebraically with separation \cite{lattice}.

\section{The single particle Green's function and condensate depletion}

    We now determine the structure of the single particle excitations in terms
of the modes of the lattice by constructing the long wavelength behavior of
the single particle Green's function,
\beq G(rt,r't') = -i\langle
  {\cal T}[(\psi(rt)-\langle\psi(rt)\rangle)(\psi^\dagger(r't')-
  \langle\psi^\dagger(r't')\rangle ]\rangle
\eeq
($\cal T$ denotes the time ordered product), in terms of the correlation
functions calculated in the previous section.
   Again expanding in small fluctuations about equilibrium, we have
\beq
  G(k,z) = \frac{1}{4\bar n}\langle nn\rangle(k,z)
     +\frac{1}{2{\bar n}}\left(\langle n e^{-i\Phi} \rangle (k,z)
      + \langle e^{i\Phi} n \rangle (k,z)\right)
     +{\bar n}\langle e^{i\Phi}e^{-i\Phi}\rangle (k,z).
\label{G}
\eeq
To second order in the fluctuations of the density and phase,
\beq
  G(k,\omega) = \frac{1}{4\bar n}\langle nn\rangle(k,z)
               +\frac i2 \left(\langle \Phi n\rangle (k,z)
                -\langle  n\Phi\rangle(k,z)\right)
                +{\bar n} \langle \Phi\Phi\rangle (k,z).
\label{G1}
\eeq

    To illustrate this method of calculating $G$, we first consider a weakly
interacting non-rotating system, for which Eqs.~(\ref{nncorr}), (\ref{D}),
(\ref{phin}), and (\ref{phiphi}) imply,
\beq
 \langle nn\rangle(k,z) &=& \frac{nk^2/m}{z^2 - E_k^2}, \\
 \langle \Phi n\rangle (k,z)& =& -\langle n\Phi\rangle (k,z) =
  -i\frac{mz}{nk^2}  \langle nn\rangle(k,z),
\eeq
and
\beq
\langle \Phi \Phi\rangle (k,z) = \frac{m^2}{k^2}\langle v_L v_L \rangle
   = \left(\frac{mz}{nk^2}\right)^2  \langle nn\rangle(k,z)
             - \frac{m}{nk^2},
\eeq
where $E_k= \left(gnk^2/m +(k^2/2m)^2\right)^{1/2}$ is the Bogoliubov
single particle energy.  (The term $\sim k^4$ in $E_k^2$ appears only if one
includes the kinetic term, $-\nabla^2n/4m$ in the pressure.)  Substitution of
these correlation functions into Eq.~(\ref{G1}) yields the usual Bogoliubov
result:
\beq
  G(k,z) = \frac{z+gn+k^2/2m}{z^2 - E_k^2}.
\eeq

    In the presence of rotation, we use Eqs.~(\ref{phiphi}) and (\ref{phin}),
to write
\beq
      G(k,z)= \frac1D\left\{(z+ms^2)\left(z^2-4\Omega^2
    -4\frac{C_1+C_2)k^2}{nm}\right)
       +\frac{k^2}{4m}\left(z^2 -2\frac{C_2}{nm}k^2 \right)\right\}.
\eeq
The contribution of the modes to the density, $n_k'$, of particles of
momentum $k$ excited out of the condensate, to leading orders in $k^2$, is
then
\beq
   n_k' = \frac{ms^2}{\omega_T}\left(f(\omega_T)+\frac12\right)
+\frac{k^2}{8m\Omega}
 \left[\left(1+\left(\frac{ms^2}{\Omega}\right)^2\right) f(\omega_I)
     +\frac12\left(\frac{ms^2}{\Omega}-1\right)^2\right].
\eeq
The contribution of the inertial mode (the final terms) is always finite
in two dimensions.  However, the contribution of the Tkachenko mode in the
soft limit (the first term) leads to a logarithmic divergence of $n'$ in two
dimensions at zero temperature ~\cite{SHM}, and a quadratic divergence at
finite temperature, thus destroying the condensation in an infinite system.

    It is instructive, finally, to calculate the effect of the lattice modes
on the superfluid mass density, $\rho_s$, which is related to the transverse
velocity autocorrelation function by \cite{standrews}
\beq
  \rho_n = \rho - \rho_s = -(mn)^2 \lim_{k\to 0}\langle v_T v_T \rangle
(k,\omega=0).
  \label{rhos}
\eeq
Since $v_T = -2\Omega\epsilon_L$, we find from Eq.~(\ref{longcorr}) that
\beq
   \rho_n = mn,
\eeq
that is, the superfluid mass density vanishes.  As discussed in
\cite{lattice} the lattice excitations replenish the sum rule (\ref{rhos}), a
reflection of the fact that the moment of inertia of a rapidly rotating
superfluid is effectively the classical value \cite{rhos}.
The vanishing of $\rho_s$ is consistent with the behavior of $G(k\to0,0)$,
which according to Josephson's sum rule should approach $-m^2n_0/\rho_s k^2$
\cite{standrews}, while in fact $G(k\to0,0) \to -(\Omega n/C_2)(2m^2/k^4)$.

\section*{Acknowledgements}

    I am grateful to S. Stringari, M. Cozzini, S.A.  Gifford, C.J.  Pethick,
V. Schweikhard, J. Anglin, and S. Vishveshwara for useful discussions.  My
thanks to the Aspen Center for Physics for hospitality during the course of
this research.  This work was supported in part by NSF Grant PHY00-98353.


\begin{references}
\bibitem{b} The interaction energy density of a rapidly rotating system is
\cite{FG} $g{\bar n}^2b/2$, where $\bar n$ is the mean density
in a unit cell of the lattice, and $b=\langle n^2\rangle/{\bar n}^2$.  The
specific value of $b$ depends on the details of the density distribution in
the cell.  At slow rotation, $b = 1 + \Omega/gn + \cdots$.  In the mean field
quantum Hall regime \cite{coresize,SHM}, $b= (e^2-5)/4(e-2)^2 = 1.158$.  We
omit the bar over the mean density when it causes no confusion.
\bibitem{sound} In a weakly interacting non-rotating homogeneous gas, $s^2
= gn/m$; however in a rotating system one must take into account the
correction $b$ to the interaction energy \cite{FG,b}.  Neglect of the
dependence of $b$ on the mean density, a reasonable approximation, implies
$s^2 = gb{\bar n}/m$, the result we use here.
\bibitem{notation} The transverse and longitudinal components of a vector
$\vec X$, e.g., $\vec\epsilon$, $\vec\zeta$, or $\vec v$, are defined by $X_T
= \vec\Omega\cdot({\hat k}\times X)$ and $X_L = {\hat k}\cdot \vec X$.
\bibitem{slinky} It should be borne in mind, however, that the vanishing
of the correlation function $\langle e^{i\Phi(r)}e^{-i\Phi(r')}\rangle$ for
large $R$ does not in itself indicate a loss of superfluidity.  Rather, the
condensate becomes fragmented.  As Josephson stressed by analogy to a very
floppy ``slinky," such vanishing correlations are not the same as the
fluctuations in the winding of the phase angle around a closed loop.
\bibitem{rhos} However, this result does not imply that the superfluid
mass density measured dynamically, e.g., in a second sound experiment,
vanishes, since in a system that is not Galilean invariant such a measurement
of $\rho_s$ is independent of a measurement via the moment of inertia.
\end{references}
\end{document}